\documentclass[utf8%,draft % раскомментировать, если нужен черновой режим
               ]{jetp} % jetp.cls
\usepackage{graphicx} % Required for inserting images
\usepackage[utf8]{inputenc}
\usepackage[T1, T2A]{fontenc}
\usepackage[english, russian]{babel}
\usepackage{latexsym}
\usepackage{amscd}
\usepackage{amsmath}
\usepackage{amssymb}
\usepackage{verbatim}
\usepackage{color}
\usepackage{indentfirst}
\usepackage{import}
\usepackage{pdfpages}
\usepackage{transparent}
\usepackage{xifthen}
\usepackage{dirtytalk}

\twocolumn % закомментировать, если не нужен двухколоночный вариант
\let\figori=\fig \def\fig[#1]#2{\figori[#1]{*}}

\newcommand{\incfig}[1]{%
    \def\svgwidth{\columnwidth}
    \import{./}{Mag_energy.png}
}
\begin{document}
\English
\title{Explosive growth of large-scale magnetic fluctuations due to particle scattering on developed small-scale Weibel turbulence in magnetoactive plasma}

\rtitle{Explosive growth of large-scale magnetic fluctuations\dots}

\author{N.~A.}{Emelyanov}\email{n.emelyanov@ipfran.ru} % a

\affiliation{A.V. Gaponov-Grekhov Institute of Applied Physics of the Russian Academy of Sciences,\\ Ulyanov str. 46, Nizhny Novgorod, 603950 Russia\\} % a

\author{Vl.~V.}{Kocharovsky}\email{kochar@ipfran.ru} % a

\affiliation{A.V. Gaponov-Grekhov Institute of Applied Physics of the Russian Academy of Sciences,\\ Ulyanov str. 46, Nizhny Novgorod, 603950 Russia\\} % a
 % a

\rauthor{N.~A.~Emelyanov, Vl.~V.~Kocharovsky}

\abstract{The analytical theory of non-linear generation of large-scale magnetic turbulence in anisotropic magnetoactive plasma in the quasilinear approximation without taking into account the direct non-linear interaction of individual harmonics is constructed. It is shown that anomalous collisions of particles due to scattering on small-scale fluctuations of the developed Weibel turbulence lead to instability of long-wave harmonics, which are stable in the linear approximation. The non-linear growth of such harmonics at a given anisotropy of the particle velocity distribution, consistent with the dynamics of short-wave perturbations at the saturation stage and possible anisotropic particle injection, occurs in the superexponential regime and corresponds to an explosive-type instability. The growth law of the large-scale magnetic field is found analytically and the critical time of explosive instability is estimated.}
\maketitle
%%%%%%%%%%%%%%%%%%%%%%%%%%%%%%%%%%%%%%%%%%%%%%%%%%%%%%%%%%%%%%%%%%%%%%
\section{Introduction}
    \label{sec1} 
%%%%%%%%%%%%%%%%%%%%%%%%%%%%%%%%%%%%%%%%%%%%%%%%%%%%%%%%%%%%%%%%%%%%%%

\par Aperiodic instability of collisionless plasma with anisotropic velocity distribution of charged particles embedded in external magnetic field has recently attracted much attention \cite{weibelSpontaneouslyGrowingTransverse1959,hamasakiElectromagneticMicroinstabilitiesPlasmas1968,landauTemperatureanisotropyInstabilityElectromagnetic1970,davidsonOrdinaryModeElectromagneticInstability1970, silvaRolePurelyTransverse2002a, tautzCounterstreamingMagnetizedPlasmas2006,lazarExistenceWeibelInstability2009,ibscherExistenceWeibelInstability2012, pokhotelovWeibelInstabilityPlasma2012,parkCollisionlessShockExperiments2015a,kocharovskyAnalyticalTheorySelfconsistent2016,grassiElectronWeibelInstability2017,emelyanovWeibelInstabilityPresence2024}. In both space and laboratory (laser) plasmas, the transition of this instability to the non-linear regime is accompanied by the development of turbulence, which essentially changes the kinetics of particles and properties of the medium. Thus, strong particle scattering, which plays the role of interparticle collisions, and anomalous viscosity, diffusion, electrical and thermal conductivity of the plasma can appear. In particular, in astrophysical plasma, the Weibel-type instability forms magnetic turbulence and largely determines the characteristics of collisionless shocks, various structures in the stellar (solar) wind, transient processes in accretion columns and discs, etc. \cite{medvedevGenerationMagneticFields1999,keenanParticleTransportRadiation2013,zhouTurbulentMixingTransition2019, zhouSpontaneousMagnetizationCollisionless2022,takabeTheoryMagneticTurbulence2023a}.
\par In many cases it is important to elucidate the properties of the magnetic turbulence in the longest wavelength part of the spectrum, including the presence of its boundary and the non-linear evolution of its energy value. A large number of works \cite{wallaceCollisionalEffectsWeibel1987, bretCharacterizationInitialFilamentation2005,schaefer-rolffsRelativisticKineticWeibel2006,stockemRelativisticFilamentationInstability2007,pokhotelovWeibelInstabilityPlasma2012,mahdaviCollisionalEffectWeibel2013a,ryutovCollisionalEffectsIon2014,aman-ur-rehmanEffectCollisionsWeibel2017, schoefflerEffectsCollisionsGeneration2020,emelyanovWeibelInstabilityPresence2024,emelyanovCollisionalMechanismExpanding2024} have been devoted to the study of the linear stage of Weibel-type instabilities for different particle velocity distribution functions, taking into account interparticle collisions and the presence of external magnetic field.  In particular, it is known that the external field in collisionless plasma stabilises Weibel instability of large-scale harmonics with wavenumbers less than a certain value $k_{min}$ and essentially allows only small-scale perturbations to grow in a limited range of wavenumbers $k_{min} < k < k_{max}$ \cite{lazarExistenceWeibelInstability2009,ibscherExistenceWeibelInstability2012,grassiElectronWeibelInstability2017,emelyanovWeibelInstabilityPresence2024}. In \cite{emelyanovCollisionalMechanismExpanding2024} it is shown that this restriction does not exist in the long wave part of the spectrum when particle collisions are taken into account. 
\par However, for the non-linear stage of the Weibel-type instability, the problem of the growing long-wavelength fluctuations remained open. Previous investigation has usually been limited to numerical modelling, which does not allow the long-term dynamics of the turbulence to be correctly followed, especially in the presence of particle collisions \cite{morseNumericalSimulationWeibel1971,davidsonNonlinearDevelopmentElectromagnetic1972,leeElectromagneticInstabilitiesFilamentation1973a,yangEvolutionWeibelInstability1994,fonsecaThreedimensionalWeibelInstability2003,borodachevDynamicsSelfConsistentMagnetic2017,takabeRecentProgressLaboratory2021,zhouSpontaneousMagnetizationCollisionless2022,takabeTheoryMagneticTurbulence2023a}. In the present work, an analytical study of the non-linear stage of Weibel-type instability has been carried out in a collisionless magnetoactive plasma, taking into account the fact that particle scattering on magnetic turbulence plays the role of anomalous collisions and can lead to the instability of long wavelength harmonics. In this case, the free energy of the anisotropic particle velocity distribution, determined by the saturation of the small-scale turbulence and possible anisotropic particle injection, is pumped into the large-scale turbulence, which has been neglected so far.
\par In sect. \ref{sec2} equations for the energy density of large- and small-scale magnetic fields are derived and solved. In sect. \ref{sec3}  the growth time of the long-wave turbulence is estimated and the peculiarities of its quasi-linear evolution are discussed. Section \ref{sec4} is the Conclusion.
%%%%%%%%%%%%%%%%%%%%%%%%%%%%%%%%%%%%%%%%%%%%%%%%%%%%%%%%%%%%%%%%%%%%%%
\section{Evolution of large- and small-scale turbulence}
\label{sec2} 
%%%%%%%%%%%%%%%%%%%%%%%%%%%%%%%%%%%%%%%%%%%%%%%%%%%%%%%%%%%%%%%%%%%%%%
\par Due to the complexity of the non-linear evolution of turbulent fields, we focus on the simplest problem of the non-relativistic Weibel instability in a plasma with stationary ions and a bi-Maxwellian distribution of electrons with concentration $n$ and anisotropy parameter $A = u_{||}^{2}/ u_{\perp}^{2} - 1 \lesssim 1$, determined by the ratio of the mean thermal velocities along $u_{||}$ and across $u_{\perp}$ (here $u_{\perp}<u_{||}$) of the external homogeneous magnetic field $\textbf{B}_0$. In this case, the magnetic energy density $B_0^2/8\pi$ is assumed to be much smaller than the electron thermal energy density $n m ( u_{\perp}^{2}+u_{||}^{2}/ 2)$, i.e. the plasma is considered to be unmagnetized. In addition, the external field should not exceed by an order of magnitude the maximum mean square field $\bar B_{s}$ of the expected turbulence \cite{borodachevDynamicsSelfConsistentMagnetic2017, emelyanovWeibelInstabilityPresence2024}. 
\par In this case, harmonics with wave vectors $\textbf{k}$ directed across the external magnetic field (and the anisotropy axis) and lying in the range between the values (see, e.g., \cite{ibscherExistenceWeibelInstability2012,emelyanovWeibelInstabilityPresence2024}): 
\begin{align}
    k_{min} \approx \frac{1+1/A}{\sqrt{2\pi}r_{H}} \ll 
k_{max} \approx\frac{\omega_{p}}{c}A^{1/2}-\frac{1+1/A}{2\sqrt{2\pi}r_{H}}
\end{align}
have the largest growth rates. Here $r_{H}=u_{\perp}/\omega_{B_{0}}$ and $\omega_{B_{0}}=e{B_{0}/mc}$ are the gyroradius and the gyrofrequency of the thermal electron with a mass $m$ and a charge $-e$, $c$ is a vacuum speed of light, $\omega_{p}=(4\pi e^{2}n/m)^{1/2}$ is a plasma frequency.
\par Outside this interval the aperiodic generation of the magnetic field without taking into account collisions of particles is actually absent, including for harmonics with tilted wave vectors corresponding to the quasi-propagating mode with non-zero real frequency, small growth rate and large projections of the electric field $\textbf{E}$ along $\textbf{k}$. 
\par Under these conditions, the well-known quasi-linear approximation \cite{achterbergWeibelInstabilityRelativistic2007,pokhotelovQuasilinearDynamicsWeibel2011,medvedevQuasinonlinearApproachWeibel2017} can be used to solve the initial Vlasov-Maxwell equations (see, e.g., \cite{kocharovskyAnalyticalTheorySelfconsistent2016}). According to it, at any moment of time, as well as at the initial linear stage of instability, the growth of the magnetic field amplitude (and current density) in each spatial harmonic occurs with the growth rate $\gamma_{k}$ determined by the dispersion relation with the value of the anisotropy parameter $A$, which is given by the concerted action of all harmonics. 
\par To simplify the demonstration of the effect of anomalous collisions caused by particle scattering on magnetic turbulence, we will assume that at the considered non-linear stage of its development the anisotropy parameter $A$ does not change significantly and is equal in magnitude to the value reached after saturation of the Weibel instability for harmonics with growth rates close to the maximum value \cite{emelyanovWeibelInstabilityPresence2024}
\begin{align}
\label{liniar_increment}
    \gamma_{max}^{(0)} =\frac{\omega_{p}}{\pi} \frac{u_{\perp}}{c}\frac{A^{3/2}}{A+1}.
\end{align}
\par The acceptability of this simplification is confirmed by numerous particle-in-cell  calculations \cite{borodachevDynamicsSelfConsistentMagnetic2017}, which show a slight change in the anisotropy parameter and a slow decay of the Weibel turbulence after the above saturation. Moreover, an approximate constancy of the anisotropy parameter can be ensured both by the injection of external electrons or by their heating by some wave fields or other (non-Weibelian) turbulence, which slightly scatters the electrons of interest.
\par At the same time, it should be taken into account that due to the appearance of anomalous collisions caused by the turbulent magnetic field, long wave harmonics with small wavenumbers $ k < k_{min} $ also become unstable. A similar model of the quasilinear description of the expanding turbulence spectrum can be constructed for other (non-bi-Maxwellian) particle distribution functions, including its evolution and the variability of the parameter $A(t)$. However, the analysis in that case becomes more complicated, in particular due to the inaccessibility of the analytical solution of the dispersion equation taking into account the magnetic field and anomalous collisions.
\par Let us introduce the spectral energy density of the turbulent magnetic field $W_{k}=\lvert B_{k}\rvert^{2}/8\pi^{3}$, which according to the above satisfies the equation
\begin{align}
    \label{inst_eq_2}
    \frac{d W_{k}}{dt}=2\gamma_{k}W_{k}.
\end{align}
\par Since in the linear approximation without collisions only short wavelength harmonics ($k_{min} < k < k_{max}$) grow in the external field and long wavelength harmonics ($0<k<k_{min}$) are stable, the spectrum naturally splits into two bands.
\par One can analyse the turbulence dynamics in the approximation of the interaction of spectral regions defined in this way, caused by the integral action of all harmonics of each band, neglecting the resonant non-linear interaction of individual modes. In this case, a change in the exact position of the boundary $k_{min}$ separating the spectral regions does not qualitatively change the result of the analysis and can be taken into account explicitly.
\par The equation (\ref{inst_eq_2}) for each of the two zones gives:   
\begin{align}
     \label{inst_syst} 
     \begin{cases}
      {\dfrac{d W_{1}}{dt_{_{_{}}}}}=2\bar{\gamma_{1}}W_{1} ,\\  
      {\dfrac{d W_{2}}{{dt}}=2\bar{\gamma_{2}}W_{2}},
     \end{cases}
    %\dfrac{d W_{1}}{dt}=2\bar{\gamma_{1}}W_{1} \nonumber\\  
     %\dfrac{d W_{2}}{dt}=2\bar{\gamma_{2}}W_{2}
\end{align}
where the following notations are introduced for the integral magnetic energy density and the integral growth rates of the corresponding spectral bands:
\begin{align}
    \label{def_energy}
    W_{1}=\int^{k_{min}}_{0}W_{k}k^{2}dk, \ \ \ W_{2}=\int^{k_{max}}_{k_{min}}W_{k}k^{2}dk, \ \ \ \\    
    \label{def_inkr}    \bar{\gamma_{1}}=\frac{\int^{k_{min}}_{0}\gamma_{k}W_{k}k^{2}dk}{\int^{k_{min}}_{0}W_{k}k^{2}dk}, \ \     \bar{\gamma_{2}}=\frac{\int^{k_{max}}_{k_{min}}\gamma_{k}W_{k}k^{2}dk}{\int^{k_{max}}_{k_{min}}W_{k}k^{2}dk}.
\end{align}

\par According to the linear theory for a collisional magnetoactive plasma\cite{emelyanovCollisionalMechanismExpanding2024}, the growth rate of the long wavelength harmonics is proportional to the electron collision frequency $\bar\nu$, i.e. $\gamma_{k}\approx\bar\nu\Phi(k)$ for $k<k_{min}$. Similarly, in this region, the growth rate also appears in the absence of collisions but in the presence of electron scattering: $\gamma_{k}=\nu_{eff}\Phi(k)$, where $\nu_{eff}$ is the effective frequency of anomalous collisions in a turbulent plasma and $\Phi(k)$ is the same form factor taken from the linear theory \cite{emelyanovCollisionalMechanismExpanding2024}, which gives a quadratic decrease of the growth rate in the region of small wavenumbers.
%$\gamma(k)$.
\par As a result,  the non-linear integral growth rate of the long-wave fluctuations can be written in the form:
%\begin{align}
%    \label{gamma_k1_val}    
%    \gamma_{k}=\nu_{eff}\Phi(k)
%\end{align}
\begin{align}
    \label{gamma1_val}    
    \bar{\gamma_{1}}=\nu_{eff}\bar{\phi}(t), \ \ \ \ \ \ \ \ \ \ \\
%\end{align}
%\begin{align}
    \label{notation_F}
    \bar{\phi}(t)=\frac{\int^{k_{min}}_{0}\Phi(k)W_{k}(t)k^{2}dk}{\int^{k_{min}}_{0}W_{k}(t)k^{2}dk}.
\end{align}
\par Of course, this only takes into account the integral effect of all short-wave fluctuations on the development of long-wave fluctuations and does not take into account the resonance interaction of individual harmonics, which could also lead to the appearance of a non-linear growth rate. However, such interaction is most significant for the fastest growing, i.e. small-scale, disturbances, but not for large-scale fluctuations. For simplicity, we also assume that the boundary $ k_{min} $ between them and the boundary of the whole instability region $ k_{max} $ are unchanged, considering the anisotropy parameter and the type of electron velocity distribution function fixed.
\par To determine the effective collision frequency $\nu_{eff}$ in the problem under consideration, it is assumed that the magnetic turbulence is isotropic and that the characteristic scale of the inhomogeneity is much smaller than the gyroradius of the particles in the external magnetic field: $\lambda_{cor}\ll r_{H}$. Then, according to\cite{batyginSovremennajaElektrodinamika2005a, keenanParticleTransportRadiation2013, medvedevQuasinonlinearApproachWeibel2017}, by order of magnitude the anomalous scattering  frequency is
\begin{align}
    \label{NU_eff}
    \nu_{eff}=\alpha\frac{e^{2}\bar B_{s}^{2}}{m^2c^2}\frac{\lambda_{cor}}{u_{th}}=\alpha\omega_{T}^2\frac{\lambda_{cor}}{u_{th}}.
\end{align} 
Here $\omega_{T}$ is the gyrofrequency of particles in the mean-square magnetic field $B_{s}$, $u_{th}$ is the mean thermal velocity of particles, $\alpha$ is a numerical multiplier of the order of one depending on the model assumptions about the properties of turbulence (in particular, its isotropy), $\lambda_{cor}$ is the correlation length of magnetic perturbations, which with accuracy to the numerical coefficient is given by the expression (see, e.g., \cite{medvedevQuasinonlinearApproachWeibel2017}): 
\begin{align}
    \label{notation_L_corr}
    {\lambda_{cor}}(t)=\frac{\int^{\infty}_{0}W_{k}(t)kdk}{\int^{\infty}_{0}W_{k}(t)k^{2}dk}.
\end{align}
In the case of $\lambda_{cor}\gtrsim r_{H}$, which is possible in particular due to the time dependence of the correlation length $\lambda_{cor}(t)$, adjustments of some of the given formulas and numerical factors are unavoidable, but the main conclusions of this paper do not change.
\par With use of the definitions (\ref{def_energy}) and (\ref{NU_eff}), one can write: 
\begin{align}
    \label{nu_val}    
    \nu_{eff}(t)=\nu_{eff}^{(s)}(t)\Big(\frac{W_{1}(t)}{W_{s}}+\frac{W_{2}(t)}{W_{s}}\Big),
\end{align}
where $\nu_{eff}^{(s)}(t)$ is the effective collision frequency calculated at the saturation energy density of magnetic turbulence $W_{s}$ and taking into account the time dependence of the correlation length (\ref{notation_L_corr}).
%$\lambda_{cor}(t)$. 
The change of thermal velocities is neglected. 
\par As can be seen from the formulas (\ref{gamma1_val}) and (\ref{nu_val}), that the large-scale harmonics growth rate of the instability already depends on the magnetic field energy in the first order of approximation, and it increases with the increasing of the turbulent fluctuations value.
\par At the stage of non-linear instability of small scale harmonics in the range $k_{min}\le k\le k_{max}$, their growth rate also depends on the turbulence level. At the initial stage of its saturation, this dependence is mainly determined by a rapid decrease of the anisotropy parameter $A$, and we do not consider this stage, since at this stage the long-wavelength disturbances of interest to us do not have the possibility or the time to grow.  The above two-zone turbulence model (\ref{inst_syst})-(\ref{def_inkr}) will be analysed starting from the stage of the appearance of a noticeable increase in long-wave fluctuations (\ref{gamma1_val}), when the level of short-wave fluctuations is already close to saturation, i.e. the value of $W_2$ approaches the value of $W_{s}$. With use of the perturbation theory, the integral growth rate $\bar\gamma_2$ can be we decomposed into a series of energy densities $W_2$: 
\begin{align}
    \label{gamma2_val}    
    \bar{\gamma_{2}}\approx \bar{\gamma_{0}}-\beta W_{2}+...
\end{align}
\par The introduced parameter $\beta$, according to the second equation of the system (\ref{inst_syst}), is related to the saturation energy density of small-scale turbulence $W_s=\bar\gamma_0/\beta$. It can be easily estimated from the equality condition of the maximum linear growth rate (\ref{liniar_increment}) and gyrofrequency in the turbulent field \cite{davidsonNonlinearDevelopmentElectromagnetic1972, bergerEquilibriumStabilityLargeAmplitude1972}: $\omega_{T}\approx\gamma_{max}^{(0)}$. For the electron Weibel instability with a bi-Maxwellian distribution function, this condition gives the following approximation of the mean-square saturation field $\bar B_{s}$ \cite{emelyanovWeibelInstabilityPresence2024}:
\begin{align}
    \label{Sat_field}    
    \frac{\bar{B}^{2}_{s}}{8\pi}\approx \frac{\omega_{p}^{2}m^2c^2}{8\pi^3e^2}\Big(\frac{u_{\perp}}{c}\Big)^2\frac{A^3}{(A+1)^2}.
\end{align}
\par By assuming $A\approx const$ and restricting the decomposition (\ref{gamma2_val}) to describe the slow growth and saturation of small-scale turbulence, we assume either its maintenance by an external source or its slow decay on the scale of the critical time of large-scale perturbations explosive instability discussed below in sect.\ref{sec3}. 
\par Introducing the dimensionless quantities $\tau=2\bar\gamma_{0}t$, $w_{1}=W_{1}/W_{s}$, $w_{2}=W_{2}/W_{s}$, $\nu=\nu_{eff}^{(s)}/\bar{\gamma_{0}}$, we write the system of equations (\ref{inst_syst}) in the following form: 
\begin{align}
    \begin{cases}
    \label{inst_syst_dim_less}
    \dfrac{d^{} w_{1}}{d\tau_{_{_{}}}}=\nu\bar{\phi}(\tau)(w_{1}+w_{2}) w_{1},\\
    \dfrac{d w_{2}}{d\tau} = w_{2}-w_{2}^{2}.
\end{cases}
\end{align}
Its integration gives the non-linear laws of saturation and growth of the magnetic field energy density of short- and long-wave Weibel turbulence, respectively:
\begin{align}
    \label{w2_integral}    
    w_{2}(\tau)=\frac{\eta \exp{(\tau)}}{1+\eta\exp(\tau)} \approx \frac{w_{2}^{(0)}\exp{(\tau)}}{1+w_{2}^{(0)}\exp(\tau)}, \ \ \ \ \ \ \ \ \ \ \ \ \  \\
%\end{align}
%\begin{align}
    \label{w1_integral}    
    w_{1}(\tau)= \ \ \ \ \ \ \ \ \ \ \ \ \ \ \ \ \ \ \ \ \ \ \ \ \ \ \ \ \ \ \ \ \ \ \ \ \ \ \ \ \ \ \ \ \ \ \ \ \ \ \ \ \ \  \  \nonumber\\ =\frac{ w_{1}^{(0)}\exp{\Big(\int_{0}^{\tau}\nu\bar\phi(\tau')w_{2}(\tau')d\tau'\Big)}}{1-w_{1}^{(0)}\int_{0}^{\tau}\nu\bar{\phi}(\tau')\exp{\Big(\int_{0}^{\tau'}\nu\bar\phi(\tau'')w_{2}(\tau'')d\tau''\Big)}d\tau'},
\end{align}
where $w_{1}^{(0)}, w_{2}^{(0)}$ are initial values for the quantities $w_{1}$ and $w_{2}$, the parameter $\eta$ is $\eta=w_{2}^{(0)}/(1-w_{2}^{(0)}) \approx w_{2}^{(0)}$. 
\par According to (\ref{w1_integral}), the growth of the mean-square magnetic field of large-scale harmonics occurs in the superexponential regime and has the character of explosive instability, since the value of $w_{1}$ formally becomes infinite in a finite time, when the denominator of the expression turns to zero. In reality, the growth of large-scale harmonics is limited by stronger non-linear effects, which are not taken into account in this two-band model and do not allow its use, when the inequality $w_{1}\gg w_{2}$ is reached. 
\par Preliminary numerical particle-in-cells calculation of the long-term dynamics of long-wavelength fluctuations in the case $w_{1} \lesssim w_{2}$ qualitatively confirms the above behaviour of the large-scale magnetic turbulence growth, and also shows its influence on the deformation of the electron velocity distribution function and the inevitable change of the anisotropy parameter $A$, which limit the applicability of the described model. Such a calculation is technically complicated, requiring a computational domain of hundreds and thousands of particle gyroradii in the external magnetic field, which corresponds to hydrodynamic scales, and therefore is beyond the scope of the present work.
%%%%%%%%%%%%%%%%%%%%%%%%%%%%%%%%%%%%%%%%%%%%%%%%%%%%%%%%%%%%%%%%%%%%%%%
\section{Critical time of explosive instability}
\label{sec3} 
%%%%%%%%%%%%%%%%%%%%%%%%%%%%%%%%%%%%%%%%%%%%%%%%%%%%%%%%%%%%%%%%%%%%%%%
\par Nevertheless, let us estimate the critical time of explosive instability $\tau_c$, determined by the condition of zero equality of the denominator of the expression (\ref{w1_integral}):
 \begin{align}
    \label{cr_time1}    
    w_{1}^{(0)}\int_{0}^{\tau_{c}}\nu\bar{\phi}(\tau')\exp{\Big(\int_{0}^{\tau'}\nu\bar\phi(\tau'')w_{2}(\tau'')d\tau''\Big)}d\tau'=1.
\end{align}
\par To obtain an explicit formula, we use the method of successive approximations, assuming that all quantities in equation (\ref{cr_time1}) are calculated at zero order, i.e. at $ w_{k} \approx w_{0k}$, where the form of the function $w_{0k}(\tau)$ is different in the two wavenumber ranges:
\begin{align}
\label{new_zero} 
   % \begin{cases}
w_{0k}(\tau)=w_{0k}^{(0)}\exp{\Big(\int_{0_{_{_{_{_{_{}}}}}}}^{\tau}\nu_{0}(\tau)\Phi(k)d\tau' \Big)}=\nonumber\\
=w_{0k}^{(0)}\exp{\big(\mu(\tau)_{_{_{_{_{_{}}}}}}\Phi(k)\big)},\ \ \  0<k<k_{min}, \nonumber\\   
    w_{0k}(\tau)=w_{0k}^{(0)}\exp{\big(\dfrac{\gamma_k}{\gamma_{0}}\tau\big)}, \ \ k_{min}\le k\le k_{max}.
    %\end{cases}
\end{align}
Here the function $\nu_{0}(\tau)$ is equal to the effective collision frequency $\nu_{eff}^{(s)}(\tau)$, for which the correlation length $\lambda_{cor}$ is taken in the zero approximation defined by the expression (\ref{new_zero}) for small wave numbers.
\par In this approximation the function (\ref{notation_F}) is:
\begin{align}
    \label{first_order_phi}   
    \bar{\phi}(\tau)\approx \frac{\int^{k_{min}}_{0}\Phi(k)\exp{\big(\mu(\tau)\Phi(k)\big)}k^{2}dk}{\int^{k_{min}}_{0}\exp{\big(\mu(\tau)\Phi(k)\big)}k^{2}dk}=
    \nonumber\\ 
    =\frac{\partial}{\partial\mu}ln\Big(\int^{k_{min}}_{0}\exp{\big(\mu\Phi(k)\big)}k^{2}dk\Big).
\end{align}
\par It follows from the linear theory for the Weibel instability in collisional magnetoactive plasma\cite{emelyanovCollisionalMechanismExpanding2024} that in the long-wavelength range ($k \lesssim k_{min}$), at not too large magnitude of anisotropy ($A\lesssim 1$) for the function $\Phi(k)$ the approximate expression $\Phi(k)\approx k^{2}l^{2}$ is valid, where $l$ is some characteristic length equal by the order of magnitude to the gyroradius of particles $r_H$ in the external magnetic field. Then, introducing the dimensionless parameter $\sigma=\mu k_{min}^{2}l^{2}$, the formula (\ref{first_order_phi}) can be written in the form:
\begin{align}
    \label{phi_integral}   
    \bar{\phi}(\sigma)\approx k_{min}^{2}l^{2}\frac{\partial}{\partial\sigma}ln\Big(\int^{1}_{0}x^{2}\exp{\big(\sigma x^{2}\big)}dx\Big),
\end{align}
where the integral under the logarithm is equal to
\begin{align}
    \label{integral}   
    \int^{1}_{0}x^{2}\exp{\big(\sigma x^{2}\big)}dx= \ \ \ \ \  \ \ \ \ \ \ \  \ \ \ \ \ \ \ \ \ \ \ \ \ \ \ \ \ \ \ \ \ \ \ \ \ \  \ \ \ \ \  \nonumber\\ =\frac{\exp{(\sigma)}}{2\sigma}\Big[1-i\sqrt{\frac{\pi}{4\sigma}}\exp{(-\sigma)\Big(1-\exp{(\sigma)}Z(\sqrt{\sigma})\Big)}\Big].
\end{align}
Here $Z(x)$ is the Faddeeva function (or Crump function). 
\par Note that the expression in square brackets is a slow (compared to the value $w_2(\tau)$) function that varies in the interval between zero and one. Therefore, the value of $\bar{\phi}$ will differ little from the constant value $\approx k_{min}^{2}l^{2}$. Similarly, it can be shown that the average collision frequency $\nu_0(\tau)$ is also a slow function and remains almost constant. Of course, the above properties will also hold for the product of these two quantities:
\begin{align}
\label{gamma_incr}
    \Gamma(\tau)=\nu_{0}(\tau)\bar{\phi}(\tau)\approx \gamma_{k}(k_{min})/\bar{\gamma_{0}},
\end{align}
where $\gamma_{k}$ corresponds to the linear growth rate in magnetoactive collisional plasma with collision frequency $\nu_{eff}^{(s)}(0)$.
\par As a result, the expression (\ref{w1_integral}) for the energy density of large-scale harmonics of the magnetic field can be written with good accuracy in the following way:
\begin{align}
    \label{w1_integral_approx_4}    
    w_{1}(\tau)\approx w_{1}^{(0)}\Big(\dfrac{1+\eta\exp{(\tau)}}{1+\eta}\Big)^{\Gamma}\Big\{1-{w_{1}^{(0)}}{}_{2}F_{1}(1+\eta)\times\nonumber\\ \times\Big[\Big(\dfrac{1+\eta\exp{(\tau)}}{1+\eta}\Big)^{\Gamma}\dfrac{_{2}F_{1}\big(1+\eta\exp{(\tau)}\big)}{_{2}F_{1}(1+\eta)}-1\Big]\Big\}^{-1}.
\end{align}
The abbreviated notation for the hypergeometric function (Gauss function) is used here $_{2}F_{1}(x)={}_{2}F_{1}(-\Gamma,1;1-\Gamma;1/x)$.
\par As a result, the equation (\ref{cr_time1}) for the critical time of explosive instability $\tau_c$ takes the form:  
\begin{align}
    \label{new_cr_time1}    
    {w_{1}^{(0)}}{}_{2}F_{1}(1+\eta)\times\ \ \ \ \ \ \ \ \ \ \ \ \ \ \ \ \ \ \ \ \ \ \ \ \ \ \ \ \ \ \ \ \ \ \ \ \ \ \ \ \ \ \ \   \ \ \ \ \  \nonumber\\ \times\Big\{  \Big(\dfrac{1+\eta\exp{(\tau_{c})}}{1+\eta}\Big)^{\Gamma}\dfrac{_{2}F_{1}(1+\eta\exp{(\tau_{c})})}{_{2}F_{1}(1+\eta)}-1\Big\}=1.
\end{align}
\par Since explosive instability requires developed small-scale turbulence, which occurs when $w_2$ reaches saturation and becomes of the order of unity, the following inequality holds for the critical time $\tau_{c}$: $\eta\exp{(\tau_{c})}\gg 1$. Then from the expression (\ref{new_cr_time1}) one can get:
\begin{align}
    \label{critical_time}    
     w_{1}^{(0)}\Big(\dfrac{\eta\exp{(\tau_{c})}}{1+\eta}\Big)^{\Gamma}\approx1.
\end{align}
\par Solving this equation, the following estimate of the critical time can be easily obtained:
\begin{align}
    \label{cr_time2}    
     \tau_{c}\approx \frac{1}{\Gamma}\ln \Big(\frac{1}{w_{1}^{(0)}w_{2}^{(0)\Gamma}}\Big).
\end{align}
\par Restricting the estimation by the order of magnitude, omitting the logarithmic multipliers, in dimensional terms the critical time for the development of explosive instability for large-scale harmonics is proportional to the inverse frequency of anomalous collisions in saturated small-scale magnetic turbulence. 
\begin{align}
    \label{cr_time3}    
     t_{c}\sim\frac{1}{\nu_{eff}^{(s)}}.
\end{align}
In other words, in a rough approximation, the critical time is equal to several free travelling times of particles in turbulent plasma. 
\par
\begin{figure}[t]  %%%%%%%%%%%%%%%%%% FIGURE 2
    \centering
    %\vspace{-0.31\textwidth} 
    \includegraphics[width=1.0\linewidth]{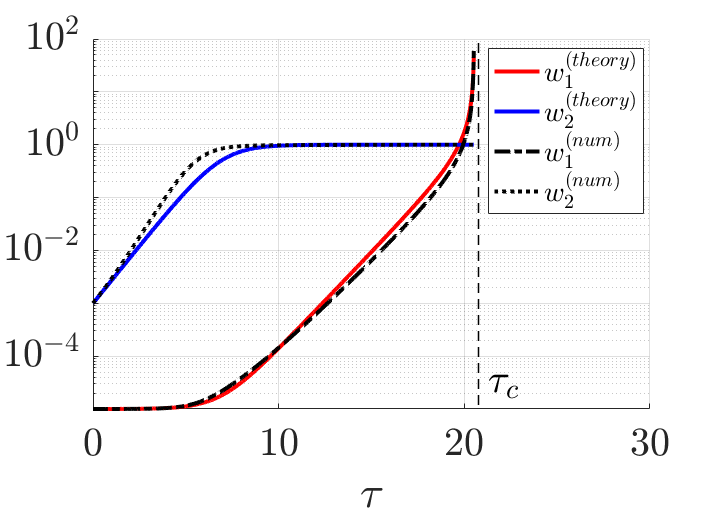} 
    
    %\scalebox{0.6}{\incfig{Mag_energy}}
    
    \caption{Time dependence of the mean-square magnetic field energy}
    \label{fig:solution}
\end{figure}

 In fig. \ref{fig:solution} the dashed and dotted curves (values of $w_{2}$ and $w_{1}$, respectively) show the results of the numerical solution of the original non-linear equation (\ref{inst_eq_2}), where the values of the growth rate $\gamma_{k}$ are taken in the quasilinear approximation, i.e, the dependence on the wave vector $\gamma({k})$ is given according to the linear theory of the Weibel instability in a magnetoactive collisional plasma \cite{emelyanovCollisionalMechanismExpanding2024} and the current value of the anomalous collision frequency and the value of the mean magnetic field are calculated in agreement with the solution for the harmonic energy $W_{k}$.  The values of the particle distribution anisotropy parameter $A=1$, the plasma parameter $\beta_{||e}=20$, and the external magnetic field $B_{0}=40$ Gs are used in the calculation.
\par For comparison, the same figure shows the analytical solutions (\ref{w2_integral}) and (\ref{w1_integral_approx_4}) (blue and red solid lines, respectively), which determine the growth law of the turbulent magnetic field. The anomalous collision frequency for saturated turbulence is $\nu_{eff}^{(s)}(0)/\omega_{p}\approx 3\cdot10^{-3}$, and the corresponding value of the quantity $\Gamma \approx 0.84$. The vertical dashed line indicates the critical time of the explosive instability.
\par The obtained analytical formulas describe well the non-linear turbulence dynamics, including near the critical time $\tau_{c}$. Its rough estimation (\ref{cr_time3}) gives the value $\tau_{c}\approx 5$, which is in order of magnitude in agreement with the more accurate result obtained from the numerical calculation and the analytical expression (\ref{new_cr_time1}). 
\par Thus, small-scale magnetic turbulence plays an essential role in the dynamics of large-scale quasi-magnetostatic perturbations in anisotropic collisionless magnetoactive plasma. The anomalous scattering of particles arising at the non-linear stage of the Weibel instability causes a superexponential growth of long-wave harmonics with a characteristic time of explosive instability of the order of several times of particles free passage in the turbulent magnetic field. 
%%%%%%%%%%%%%%%%%%%%%%%%%%%%%%%%%%%%%%%%%%%%%%%%%%%%%%%%%%%%%%%%%%%%%%%
\vspace{-4mm}
\section{Conclusion}
\label{sec4} 
\vspace{-2mm}
%%%%%%%%%%%%%%%%%%%%%%%%%%%%%%%%%%%%%%%%%%%%%%%%%%%%%%%%%%%%%%%%%%%%%%%
\par The formulated model of large-scale turbulence dynamics, the analytical solutions obtained and the qualitative conclusions drawn can be extended to more general cases of Weibel-type instability development in multi-component, relativistic, weakly collisional magnetoactive plasma with a wide class of particle velocity distribution functions \cite{schaefer-rolffsRelativisticKineticWeibel2006,kocharovskyAnalyticalTheorySelfconsistent2016}. The main conditions are i) the applicability of the quasilinear approximation, i.e. the absence of strongly localised non-linear structures and effective non-linear interaction of separate turbulence harmonics, and ii) the correctness of  separation in the spectrum of the long- and short-wave resulting turbulence, in which there is no or a significant instability growth rate before turbulence excitation, respectively.
\par Under such conditions, according to the above, one should expect at long-term preservation and significant value of anisotropy of particle velocity distribution after saturation of small-scale turbulence or in the presence of effective anisotropic injection of particles (for example, due to any wave heating or inflow of external matter):
\par (a) non-linear instability of large-scale magnetic field fluctuations with wave numbers smaller than the minimum value for the Weibel instability in a given magnetic field, arising due to particle scattering on the forming small-scale turbulence,
\par (b) increase of the growth rate of this instability due to strengthening of anomalous collisions at the expense of additional scattering of particles on the arising large-scale turbulence and superexponential (explosive) growth of its magnetic energy up to the level of the order or even more than the energy of saturated small-scale turbulence,
\par (c) stopping of the specified explosive process after a time of the order of several free path times of non-magnetised charged particles in the plasma with saturated magnetic turbulence.
\par The proposed model and the general scenario of non-linear evolution of large-scale magnetic turbulence need to be refined and detailed for each specific situation in collisionless magnetoactive plasma.
\par Strictly speaking, despite the good agreement of the model conclusions with numerical calculations for the simplest problem considered in sect.\ref{sec2} of the present work, the following questions remain open:
\par (1) the role of the non-isotropy of the turbulent spectrum and the resonance interaction of its harmonics, 
\par (2) the influence of a concerted change in the anisotropy of the particle velocity distribution under the action of a changing turbulent field and the process of inevitable saturation of the explosive instability, 
\par (3) the change of the correlation length of the random magnetic field and rearrangement of the spectrum of large-scale turbulence in the course of its development, and etc. 
\par Nevertheless, the developed mechanism of non-linear formation of the large-scale magnetic perturbations seems to be useful and promising for solving a number of problems and interpreting observations in astrophysical and laboratory weakly collisional plasmas. Examples are multiscale quasimagnetostatic electric current structures in the solar corona, especially in its active regions and during solar flares and coronal mass ejections.
~\\
\par The work was carried out under a grant from the Theoretical Physics and Mathematics Advancement Foundation “BASIS” No 24-1-1-97-5.
%%% BIBLIOGRAPHY %%%%%%%%%%%%%%%%%%%%%%%%%%%%%%%%%%%%%%%%%%%%%%%

\bibliography{biblio_JETP}

\begin{thebibliography}{10}

\bibitem{weibelSpontaneouslyGrowingTransverse1959}
E.~S. Weibel, ``Spontaneously {{Growing Transverse Waves}} in a {{Plasma Due}} to an {{Anisotropic Velocity Distribution}},'' {\em Physical Review Letters}, vol.~2, no.~3, pp.~83--84, 1959-02-01.

\bibitem{hamasakiElectromagneticMicroinstabilitiesPlasmas1968}
S.~Hamasaki, ``Electromagnetic {{Microinstabilities}} of {{Plasmas}} in a {{Uniform Magnetic Induction}},'' {\em The Physics of Fluids}, vol.~11, no.~12, pp.~2724--2727, 1968-12-01.

\bibitem{landauTemperatureanisotropyInstabilityElectromagnetic1970}
R.~W. Landau and S.~Cuperman, ``A temperature-anisotropy instability for electromagnetic waves propagating across a static magnetic field,'' {\em Journal of Plasma Physics}, vol.~4, no.~1, pp.~13--20, 1970-02.

\bibitem{davidsonOrdinaryModeElectromagneticInstability1970}
R.~C. Davidson and C.~S. Wu, ``Ordinary-{{Mode Electromagnetic Instability}} in {{High-$\beta$ Plasmas}},'' {\em The Physics of Fluids}, vol.~13, no.~5, pp.~1407--1409, 1970-05-01.

\bibitem{silvaRolePurelyTransverse2002a}
L.~O. Silva, R.~A. Fonseca, J.~W. Tonge, W.~B. Mori, and J.~M. Dawson, ``On the role of the purely transverse {{Weibel}} instability in fast ignitor scenarios,'' {\em Physics of Plasmas}, vol.~9, no.~6, pp.~2458--2461, 2002-06-01.

\bibitem{tautzCounterstreamingMagnetizedPlasmas2006}
R.~C. Tautz and R.~Schlickeiser, ``Counterstreaming magnetized plasmas. {{II}}. {{Perpendicular}} wave propagation,'' {\em Physics of Plasmas}, vol.~13, no.~6, p.~062901, 2006-06-01.

\bibitem{lazarExistenceWeibelInstability2009}
M.~Lazar, R.~Schlickeiser, and S.~Poedts, ``On the existence of {{Weibel}} instability in a magnetized plasma. {{I}}. {{Parallel}} wave propagation,'' {\em Physics of Plasmas}, vol.~16, no.~1, p.~012106, 2009-01-01.

\bibitem{ibscherExistenceWeibelInstability2012}
D.~Ibscher, M.~Lazar, and R.~Schlickeiser, ``On the existence of {{Weibel}} instability in a magnetized plasma. {{II}}. {{Perpendicular}} wave propagation: {{The}} ordinary mode,'' {\em Physics of Plasmas}, vol.~19, no.~7, p.~072116, 2012-07-01.

\bibitem{pokhotelovWeibelInstabilityPlasma2012}
O.~A. Pokhotelov and M.~A. Balikhin, ``Weibel instability in a plasma with nonzero external magnetic field,'' {\em Annales Geophysicae}, vol.~30, no.~7, pp.~1051--1054, 2012-07-13.

\bibitem{parkCollisionlessShockExperiments2015a}
H.~S. Park, C.~M. Huntington, F.~Fiuza, R.~P. Drake, D.~H. Froula, G.~Gregori, and et~al., ``Collisionless shock experiments with lasers and observation of {{Weibel}} instabilities,'' {\em Physics of Plasmas}, vol.~22, no.~5, p.~056311, 2015-05-01.

\bibitem{kocharovskyAnalyticalTheorySelfconsistent2016}
V.~V. Kocharovsky, V.~V. Kocharovsky, V.~J. Martyanov, and S.~V. Tarasov, ``The analytical theory of self-consistent current structures in a collisionless plasma,'' {\em Uspekhi Fizicheskih Nauk}, vol.~186, no.~12, pp.~1267--1314, 2016-12.

\bibitem{grassiElectronWeibelInstability2017}
A.~Grassi, M.~Grech, F.~Amiranoff, F.~Pegoraro, A.~Macchi, and C.~Riconda, ``Electron {{Weibel}} instability in relativistic counterstreaming plasmas with flow-aligned external magnetic fields,'' {\em Physical Review E}, vol.~95, no.~2, p.~023203, 2017-02-03.

\bibitem{emelyanovWeibelInstabilityPresence2024}
N.~A. Emelyanov and V.~V. Kocharovsky, ``Weibel {{Instability}} in the {{Presence}} of an {{External Magnetic Field}}: {{Analytical Results}},'' {\em Radiophysics and Quantum Electronics}, vol.~66, no.~9, pp.~664--678, 2024-02.

\bibitem{medvedevGenerationMagneticFields1999}
M.~V. Medvedev and A.~Loeb, ``Generation of {{Magnetic Fields}} in the {{Relativistic Shock}} of {{Gamma-Ray-Burst Sources}},'' {\em The Astrophysical Journal}, vol.~526, no.~2, pp.~697--706, 1999-12.

\bibitem{keenanParticleTransportRadiation2013}
B.~D. Keenan and M.~V. Medvedev, ``Particle transport and radiation production in sub-{{Larmor-scale}} electromagnetic turbulence,'' {\em Physical Review E}, vol.~88, no.~1, p.~013103, 2013-07-12.

\bibitem{zhouTurbulentMixingTransition2019}
Y.~Zhou, T.~T. Clark, D.~S. Clark, S.~Gail~Glendinning, M.~Aaron~Skinner, C.~M. Huntington, O.~A. Hurricane, A.~M. Dimits, and B.~A. Remington, ``Turbulent mixing and transition criteria of flows induced by hydrodynamic instabilities,'' {\em Physics of Plasmas}, vol.~26, no.~8, p.~080901, 2019-08-01.

\bibitem{zhouSpontaneousMagnetizationCollisionless2022}
M.~Zhou, V.~Zhdankin, M.~W. Kunz, N.~F. Loureiro, and D.~A. Uzdensky, ``Spontaneous magnetization of collisionless plasma,'' {\em Proceedings of the National Academy of Sciences}, vol.~119, no.~19, p.~e2119831119, 2022-05-10.

\bibitem{takabeTheoryMagneticTurbulence2023a}
H.~Takabe, ``Theory of magnetic turbulence and shock formation induced by a collisionless plasma instability,'' {\em Physics of Plasmas}, vol.~30, no.~3, p.~030901, 2023-03-01.

\bibitem{wallaceCollisionalEffectsWeibel1987}
J.~M. Wallace, J.~U. Brackbill, C.~W. Cranfill, D.~W. Forslund, and R.~J. Mason, ``Collisional effects on the {{Weibel}} instability,'' {\em The Physics of Fluids}, vol.~30, no.~4, pp.~1085--1088, 1987-04-01.

\bibitem{bretCharacterizationInitialFilamentation2005}
A.~Bret, M.-C. Firpo, and C.~Deutsch, ``Characterization of the {{Initial Filamentation}} of a {{Relativistic Electron Beam Passing}} through a {{Plasma}},'' {\em Physical Review Letters}, vol.~94, no.~11, p.~115002, 2005-03-22.

\bibitem{schaefer-rolffsRelativisticKineticWeibel2006}
U.~Schaefer-Rolffs, I.~Lerche, and R.~Schlickeiser, ``The relativistic kinetic {{Weibel}} instability: {{General}} arguments and specific illustrations,'' {\em Physics of Plasmas}, vol.~13, no.~1, p.~012107, 2006-01-01.

\bibitem{stockemRelativisticFilamentationInstability2007}
A.~Stockem, I.~Lerche, and R.~Schlickeiser, ``The {{Relativistic Filamentation Instability}} in {{Magnetized Plasmas}},'' {\em The Astrophysical Journal}, vol.~659, no.~1, pp.~419--425, 2007-04-10.

\bibitem{mahdaviCollisionalEffectWeibel2013a}
M.~Mahdavi and H.~Khanzadeh, ``Collisional effect on the {{Weibel}} instability with the bi-{{Maxwellian}} distribution function,'' {\em Physics of Plasmas}, vol.~20, no.~5, p.~052114, 2013-05-01.

\bibitem{ryutovCollisionalEffectsIon2014}
D.~D. Ryutov, F.~Fiuza, C.~M. Huntington, J.~S. Ross, and H.-S. Park, ``Collisional effects in the ion {{Weibel}} instability for two counter-propagating plasma streams,'' {\em Physics of Plasmas}, vol.~21, no.~3, p.~032701, 2014-03-01.

\bibitem{aman-ur-rehmanEffectCollisionsWeibel2017}
{Aman-ur-Rehman}, S.~Ali~Shan, and T.~Majeed, ``Effect of collisions on {{Weibel}} instability with anisotropic electron distributions,'' {\em Physics of Plasmas}, vol.~24, no.~12, p.~122113, 2017-12-01.

\bibitem{schoefflerEffectsCollisionsGeneration2020}
K.~M. Schoeffler and L.~O. Silva, ``Effects of collisions on the generation and suppression of temperature anisotropies and the {{Weibel}} instability,'' {\em Physical Review Research}, vol.~2, no.~3, p.~033233, 2020-08-11.

\bibitem{emelyanovCollisionalMechanismExpanding2024}
N.~A. Emelyanov and V.~V. Kocharovsky, ``Collisional {{Mechanism}} of {{Expanding Wavenumbers Range}} of {{Weibel-Type Instability}} in {{Magnetoactive Plasma}},'' {\em Plasma Physics Reports}, vol.~50, no.~2, pp.~199--205, 2024-02.

\bibitem{morseNumericalSimulationWeibel1971}
R.~L. Morse and C.~W. Nielson, ``Numerical {{Simulation}} of the {{Weibel Instability}} in {{One}} and {{Two Dimensions}},'' {\em The Physics of Fluids}, vol.~14, no.~4, pp.~830--840, 1971-04-01.

\bibitem{davidsonNonlinearDevelopmentElectromagnetic1972}
R.~C. Davidson, D.~A. Hammer, I.~Haber, and C.~E. Wagner, ``Nonlinear {{Development}} of {{Electromagnetic Instabilities}} in {{Anisotropic Plasmas}},'' {\em The Physics of Fluids}, vol.~15, no.~2, pp.~317--333, 1972-02-01.

\bibitem{leeElectromagneticInstabilitiesFilamentation1973a}
R.~Lee and M.~Lampe, ``Electromagnetic {{Instabilities}}, {{Filamentation}}, and {{Focusing}} of {{Relativistic Electron Beams}},'' {\em Physical Review Letters}, vol.~31, no.~23, pp.~1390--1393, 1973-12-03.

\bibitem{yangEvolutionWeibelInstability1994}
T.-Y.~B. Yang, J.~Arons, and A.~B. Langdon, ``Evolution of the {{Weibel}} instability in relativistically hot electron–positron plasmas,'' {\em Physics of Plasmas}, vol.~1, no.~9, pp.~3059--3077, 1994-09-01.

\bibitem{fonsecaThreedimensionalWeibelInstability2003}
R.~A. Fonseca, L.~O. Silva, J.~W. Tonge, W.~B. Mori, and J.~M. Dawson, ``Three-dimensional {{Weibel}} instability in astrophysical scenarios,'' {\em Physics of Plasmas}, vol.~10, no.~5, pp.~1979--1984, 2003-05-01.

\bibitem{borodachevDynamicsSelfConsistentMagnetic2017}
L.~Borodachev, M.~Garasev, D.~O. Kolomiets, V.~V. Kocharovsky, g.-i. Martyanov, V.Yu., and A.~Nechaev, ``Dynamics of a {{Self-Consistent Magnetic Field}} and {{Diffusive Scattering}} of {{Ions}} in a {{Plasma}} with {{Strong Thermal Anisotropy}},'' {\em Radiophysics and Quantum Electronics}, vol.~59, no.~12, pp.~991--999, 2017-05-01.

\bibitem{takabeRecentProgressLaboratory2021}
H.~Takabe and Y.~Kuramitsu, ``Recent progress of laboratory astrophysics with intense lasers,'' {\em High Power Laser Science and Engineering}, vol.~9, p.~e49, 2021.

\bibitem{achterbergWeibelInstabilityRelativistic2007}
A.~Achterberg, J.~Wiersma, and C.~A. Norman, ``The {{Weibel}} instability in relativistic plasmas: {{II}}. {{Nonlinear}} theory and stabilization mechanism,'' {\em Astronomy \& Astrophysics}, vol.~475, no.~1, pp.~19--36, 2007-11.

\bibitem{pokhotelovQuasilinearDynamicsWeibel2011}
O.~A. Pokhotelov and O.~A. Amariutei, ``Quasi-linear dynamics of {{Weibel}} instability,'' {\em Annales Geophysicae}, vol.~29, no.~11, pp.~1997--2001, 2011-11-04.

\bibitem{medvedevQuasinonlinearApproachWeibel2017}
M.~V. Medvedev, ``Quasi-nonlinear approach to the {{Weibel}} instability,'' {\em [arXiv]}, 2017-05-09.

\bibitem{batyginSovremennajaElektrodinamika2005a}
V.~V. Batygin and I.~N. Toptygin, {\em Sovremennaja elektrodinamika}.
\newblock Inst. Kompjuternych Issledovanij [u.a.], 2005.

\bibitem{bergerEquilibriumStabilityLargeAmplitude1972}
R.~L. Berger and R.~C. Davidson, ``Equilibrium and {{Stability}} of {{Large-Amplitude Magnetic Bernstein-Greene-Kruskal Waves}},'' {\em The Physics of Fluids}, vol.~15, no.~12, pp.~2327--2340, 1972-12-01.

\end{thebibliography}
%%%%%%%%%%%%%%%%%%%%%%%%%%%%%%%%%%%%%%%%%%%%%%%%%%%%%%%%%%%%%%%%

\end{document}